\documentclass{article}
\usepackage{amsfonts}

\newtheorem{theorem}{Theorem}
\newtheorem{lemma}[theorem]{Lemma}

\newtheorem{conjecture}[theorem]{Conjecture}

\newtheorem{definition}[theorem]{Definition}
\newtheorem{corollary}[theorem]{Corollary}

\newtheorem{remark}{Remark}
\newtheorem{condition}[theorem]{Condition}
\def\stackunder#1#2{\mathrel{\mathop{#2}\limits_{#1}}}
\begin{document}

\title{ON ABSENCE OF EMBEDDED EIGENVALUES FOR SCHR\~{O}DINGER OPERATORS WITH
PERTURBED PERIODIC POTENTIALS}
\author{Peter Kuchment \\
Department of Mathematics and Statistics\\
Wichita State University\\
Wichita, KS 67260-0033\\
kuchment@twsuvm.uc.twsu.edu\\
http://www.math.twsu.edu/Faculty/Kuchment/ \and Boris Vainberg \\
Mathematics Department\\
University of North Carolina\\
Charlotte, NC 28223\\
brvainbe@uncc.edu}
\date{}
\maketitle

\begin{abstract}
The problem of absence of eigenvalues imbedded into the continuous spectrum
is considered for a Schr\"{o}dinger operator with a periodic potential
perturbed by a sufficiently fast decaying ``impurity'' potential. Results of
this type have previously been known for the one-dimensional case only.
Absence of embedded eigenvalues is shown in dimensions two and three if the
corresponding Fermi surface is irreducible modulo natural symmetries. It is
conjectured that all periodic potentials satisfy this condition. Separable
periodic potentials satisfy it, and hence in dimensions two and three
Schr\"{o}dinger operator with a separable periodic potential perturbed by a
sufficiently fast decaying ``impurity'' potential has no embedded
eigenvalues.

\textbf{1991 Subject Classification: }35P99, 47A55, 47F05

\textbf{Keywords}: Schr\"{o}dinger operator, periodic potential, embedded
eigenvalues
\end{abstract}

\section{Introduction}

Consider the stationary Schr\"{o}dinger operator 
\begin{equation}
H_0=-\Delta +q(x)  \label{H0}
\end{equation}
in $L^2(\bf{R}^n)$ ($n=2$ or $3$), where the real potential $q(x)\in
L^\infty (\bf{R}^n)$ is periodic with respect to the integer lattice $\bf{Z%
}^n$: $q(x+l)=q(x)$ for all $l\in \bf{Z}^n$, $x\in \bf{R}^n$. The spectrum
of this operator is absolutely continuous and has the well known band-gap
structure (see \cite{E}, \cite{Gelf}, \cite{K}, \cite{OK}, \cite{RS}, \cite
{T}, \cite{W}): 
\[
\sigma (H_0)=\stackunder{i\geq 1}{\cup }[a_i,b_i],
\]
where $a_i<b_i$, and $\lim\nolimits_{i\rightarrow \infty }a_i=\infty $. Let
us introduce also a perturbed operator with an impurity potential $v(x)$: 
\begin{equation}
H=H_0+v(x)=-\Delta +q(x)+v(x),  \label{H}
\end{equation}
where $v(x)$ is compactly supported or sufficiently fast decaying at
infinity (the exact conditions will be introduced later). It is known (see 
\cite{B2}, Section 18 in \cite{G}, and references therein) that the
continuous spectrum of $H$ coincides with the spectrum of $H_0$, and only
some additional ``impurity'' point spectrum $\{\lambda _j\}$ can arise. A
general understanding is that eigenvalues $\lambda _j$ should normally arise
only in the gaps of the continuous spectrum (i.e. in the gaps of $\sigma
(H_0)$). In other words, the case of embedded eigenvalues when one of the
eigenvalues $\lambda _j$ belongs to one of the open segments $(a_i,b_i)$ is
prohibited (at least for a sufficiently fast decaying impurity potential $%
v(x)$). Physically speaking, the existence of an embedded eigenvalue means a
strange situation when an electron is confined, in spite of having enough
energy for propagation. There are known examples of embedded eigenvalues,
first of which was suggested by J. von Neumann and E. Wigner (see \cite{EK}
and section XIII.13 in \cite{RS} for discussion of this topic and related
references). The problem of the absence of embedded eigenvalues has been
intensively studied. There are many known results on non-existence of
embedded eigenvalues for the case of zero underlying potential $q(x)$ (see,
for instance, books \cite{EK}, \cite{H} Section 14.7, and \cite{RS} Section
XIII.13). The case of a periodic potential $q(x)$ is much less studied. In 
\cite{EK} a one-dimensional example is provided where a not very fast
decaying perturbation of a periodic potential creates embedded eigenvalues.
There are many papers devoted to studying the behavior of the point spectrum
in the gaps of the continuous spectrum (see, for instance, \cite{AADH} - 
\cite{B3}, \cite{D}, \cite{GS}, \cite{He}, \cite{L}-\cite{L3}, \cite{Ra},
and \cite{RB} - \cite{S}). However, apparently the only known result on the
absence of embedded eigenvalues relates to the one-dimensional case of the
Hill's operator 
\[
H_0=-\frac{d^2}{dx^2}+q(x)+\nu (x).
\]
(see \cite{RB}, \cite{RB2}). The purpose of this paper is to address the
multi-dimensional case. Section 2 contains some necessary notions, auxiliary
information, and our main condition on the periodic potential. It is
conjectured that the condition is always satisfied. According to \cite{BKT},
this condition is satisfied at least when the potential is separable in 2D
or of the form $q_1(x_1)+q_2(x_2,x_3)$ in 3D. In Section 3 we prove a
conditional statement (in dimension less than four) on the absence of
embedded eigenvalues for a periodic potential satisfying this condition.
Finally, the Section 4 contains the main unconditional result. In our
considerations we follow an approach that was suggested long ago by the
second author for the case of zero underlying potential $q(x)$ (see \cite{V}%
). Some recent developments made it possible to adjust this method to the
periodic case. In particular, results on allowed rate of decay of solutions
obtained in \cite{FH} and \cite{M} play a crucial role.

A more limited version of the main result was announced by the authors
without proof in \cite{KV}.

\section{\label{Bloch}Bloch and Fermi varieties}

Let us introduce some notions and notations. We assume that $q(x)\in
L^\infty (\bf{R}^n)$ is a periodic potential.

\begin{definition}
The (complex) \textbf{Bloch variety} $B(q)$ of the potential $q$ consists of
all pairs $(k,\lambda )\in \bf{C}^{n+1}$ ( where $k\in \bf{C}^n$ is a 
\textbf{quasimomentum} and $\lambda \in \bf{C}$ is an eigenvalue) for which
there exists a non-zero solution of the equation 
\begin{equation}
H_0u=\lambda u  \label{spect_0}
\end{equation}
satisfying the so called Floquet-Bloch condition: 
\begin{equation}
u(x+l)=e^{ik\cdot l}u(x),\;l\in \bf{Z}^n,\;x\in \bf{R}^n.  \label{Fl}
\end{equation}
Here $k\cdot l=\sum k_il_i$ is the standard dot-product.
\end{definition}

\begin{definition}
The projection $F_c(q)$ onto $\bf{C}^n$ of the intersection of the Bloch
variety $B(q)$ with the subspace $\lambda =c$ is called the (complex) 
\textbf{Fermi variety}\textit{\ of} the potential $q$ at the level of energy 
$c$: 
\[
F_c(q)=\left\{ k\in \bf{C}^n|\;(k,c)\in B(q)\right\} .
\]
In other words, $F_c(q)$ consists of all quasimomenta $k\in \bf{C}^n$ for
which there exists a non-zero solution of the equation 
\[
H_0u=cu
\]
satisfying (\ref{Fl}).
\end{definition}

When $c=0$ we will use the notation $F(q)$ for $F_0(q)$. Let us notice the
following obvious relation between the Bloch and Fermi varieties: 
\begin{equation}
\left\{ (k,c)\in B(q)\right\} \Longleftrightarrow \left\{ k\in
F_c(q)\right\} \Longleftrightarrow \left\{ k\in F(q-c)\right\} .
\label{equiv}
\end{equation}

We will use the following notations for the real parts of the above
varieties: 
\[
B_{\bf{R}}(q)=B(q)\cap \bf{R}^{n+1},F_{\bf{R},\bf{\lambda }%
}(q)=F_{\,\lambda }(q)\cap \bf{R}^n. 
\]
The varieties $B_{\bf{R}}(q)$ and $F_{\bf{R},\lambda }(q)$ are called the 
\textbf{real Bloch variety} and the \textbf{real Fermi variety}
respectively. The following statement is contained in Theorems 3.17 and
4.4.2 of \cite{K} (with remarks of Section 3.4.D in \cite{K} about
conditions on potentials taken into account).

\begin{lemma}
\label{analyt}The Bloch variety $B(q)\subset \bf{C}^{n+1}$ is the set of
all zeros of a non-zero entire function $f(k,\lambda )$ on $\bf{C}^{n+1}$
of order $n$ (see \cite{Lel}), i.e. 
\[
\left| f(k,\lambda )\right| \leq C_{\,\epsilon }\exp \left( (\left| k\right|
+\left| \lambda \right| )^{n+\epsilon }\right) ,\;\forall \;\epsilon >0
.
\]
The Fermi variety $F_{\,\lambda }(q)$ is the set of all zeros of a non-zero
entire function of order $n$ on $\bf{C}^n$.
\end{lemma}

Lemma \ref{analyt} implies in particular that both Bloch and Fermi varieties
are examples of what is called in complex analysis \textbf{analytic sets}
(see for instance \cite{Ch}, \cite{GR}, and \cite{N}). Moreover, these are 
\textbf{principal} analytic sets in the sense that they are sets of all
zeros of single analytic functions, while general analytic sets might
require several analytic equations for their (local) description.
Analyticity of these varieties (without estimates on the grows of the
defining function) was obtained in \cite{W}.

\begin{definition}
An analytic set $A\subset \bf{C}^m$ is said to be \textbf{irreducible}, if
it cannot be represented as the union of two proper analytic subsets.
\end{definition}

Irreducibility of the zero set of an analytic function can be understood as
absence of non-trivial factorizations of this function (i.e., of a
factorization into analytic factors that have smaller zero sets).

\begin{definition}
A point of an analytic set $A\subset \bf{C}^m$ is said to be \textbf{regular%
}, if in a neighborhood of this point the set $A$ can be represented as an
analytic submanifold of $\bf{C}^m$. The set of all regular points of $A$ is
denoted by $regA$.
\end{definition}

We collect in the following lemma several basic facts about analytic sets
that we will need later. The reader can find them in many books on several
complex variables. In particular, all these statements are proven in
sections 2.3, 5.3, 5.4, and 5.5 of Chapter 1 of \cite{Ch}.

\begin{lemma}
\label{property}Let $A$ be an analytic set.

a) The set $regA$ is dense in $A$. Its complement in $A$ is closed and
nowhere dense in $A$.

b) The set $A$ can be represented as a (maybe infinite) locally finite union
of irreducible subsets 
\[
A=\stackunder{i}{\cup }A_i
\]
called its \textbf{irreducible components}.

c) Irreducible components are closures of connected components of $regA$. In
particular, set $A$ is irreducible if and only if $regA$ is connected.

d) Let $A$ be irreducible and $A_1$ be another analytic set such that $A\cap
A_1$ contains a non-empty open portion of $A$. Then $A\subset A_1$. In
particular, if $f$ is an analytic function that vanishes on an open portion
of $A$, then $f$ vanishes on $A$.

e) Any analytic set $A$ has a stratification $A=\cup A_j$ into disjoint
complex analytic manifolds (strata) $A_j$ such that the union $\cup A_j$ is
locally finite, the closure $\overline{A_j}$ of each $A_j$ and its boundary $%
\overline{A_j}\backslash A_j$ are analytic subsets, and such that if the
intersection $A_j\cap \overline{A_k}$ of two different strata is not empty,
then $A_j\subset \overline{A_k}$ and $dimA_j<dimA_k$.
\end{lemma}

We will need the following simple corollary from this lemma.

\begin{corollary}
\label{cor}Let $A$ be an irreducible proper analytic subset of $\bf{C}^n$
such that the intersection $A\cap \bf{R}^n$ contains an open part of a
smooth $(n-1)$-dimensional submanifold $M\subset \bf{R}^n$. If $f$ is an
analytic function in a complex neighborhood of $M$ such that $f=0$ on $M$,
then $f=0$ on an open subset of $A$. In particular, if $B$ is another
analytic subset of $\bf{C}^n$ such that $M\subset B$, then $A\subset B$.
\end{corollary}

\textbf{Proof}. Considering stratification $A=\cup A_j$ (see the last
statement in the preceding lemma), one can conclude that only strata of
dimension $n-1$ can contain open pieces of $\dot{M}$. In fact, if there is
an $A_j$ of complex dimension at most $n-2$ containing an open piece of $M$,
we get the contradiction as follows. Consider a point of $M$ and the tangent
space to $A_j$ at this point. This is a complex linear subspace of complex
dimension at most $n-2$, which contains a real subspace (tangent space to $M$%
) of real dimension $n-1$. This is obviously impossible. So, now we can
assume that instead of $A$ we are dealing with one of its strata of
dimension $n-1$, i.e. we may assume that $A$ is smooth. In appropriate
analytic coordinates in a complex neighborhood of a point of $M$ one can
represent $A$ locally as a complex hyperplane with the real part $M$. Then
standard uniqueness theorem for analytic functions implies that any function 
$f$ analytic on $A$ and vanishing on $M$ is identically equal to zero on $A$%
. If now $B$ is an analytic subset of $\bf{C}^n$ such that $M\subset B$,
then each of the analytic functions locally defining $B$ has the properties
of the function $f$ above. Hence, we conclude that $B$ contains an open part
of $A$. Then statement d) of the lemma implies that $A\subset B$. This
finishes the proof of the corollary.

After this brief excursion into complex analysis we return now to Bloch and
Fermi varieties. The next statement follows by inspection of (\ref{Fl}):

\begin{lemma}
The sets $B(q)$ and $F_{\,\lambda }(q)$ are periodic with respect to the
quasimomentum $k$ with the lattice of periods $2\pi \bf{Z}^n\subset \bf{C}%
^n$ (i.e., the dual lattice to $\bf{Z}^n$).
\end{lemma}

We choose as a fundamental domain of the group $2\pi \bf{Z}^n$ acting on $%
\bf{R}^n$ the following set called the (first) \textbf{Brillouin zone}: 
\[
B=\left\{ k=(k_1,...,k_n)\in \bf{R}^n|\;0\leq k_j\leq 2\pi
,\;j=1,...,n\right\} .
\]

It has been known for a long time (though not always formulated in these
terms) that one can tell where the spectrum $\sigma (H_0)$ lies by looking
at the Fermi variety. We will now briefly remind the reader this classical
result and, at the same time, the definition of spectral bands.

Consider the problem (\ref{spect_0}) - (\ref{Fl}). It has a discrete real
spectrum $\{\lambda _j(k)\}$, where $\lambda _j\rightarrow \infty $ as $%
j\rightarrow \infty $. We number the eigenvalues in the increasing order.
This way we obtain a sequence of continuous functions $\lambda _j(k)$ on the
Brillouin zone $B$. They are usually called \textbf{band functions}, or
branches of the \textbf{dispersion relation}. The values of the function $%
\lambda _j(k)$ for a fixed $j$ span the $j$th band $[a_j,b_j]$ of the
spectrum $\sigma (H_0)$. A reformulation of this statement is the following
theorem, which can be found in equivalent forms in \cite{E} (Section 6.6), 
\cite{Gelf}, \cite{K} (Theorem 4.1.1), \cite{OK}, and \cite{RS} (Theorem
XII.98).

\begin{theorem}
A point $\lambda $ belongs to the spectrum $\sigma (H_0)$ of the operator $%
H_0=-\Delta +q(x)$ if and only if the real Fermi variety $F_{\bf{R},\lambda
}(q)$ is non-empty.
\end{theorem}

In other words, by changing $\lambda $ one observes the Fermi variety $%
F_{\,\lambda }(q)$ and notices the moments when it touches the real
subspace. This set of values of $\lambda $ is the spectrum. Now the question
arises how can one distinguish the interiors of the spectral bands. The
natural idea is that when $\lambda $ is in the interior of a spectral band,
then the real Fermi variety will be massive. ``Massive'' means here ``of
dimension $n-1$'', i.e. of maximal possible dimension for a proper analytic
subset in $\bf{R}^n$. This is confirmed by the following statement.

\begin{lemma}
\label{big}If $\lambda $ belongs to the interior of a spectral band, then
the real Fermi variety $F_{\bf{R},\lambda }(q)$ contains an open piece of a
submanifold $M$ of dimension $n-1$ in $\bf{R}^n$.
\end{lemma}

Proof of the lemma follows from the stratification of the real Fermi variety 
$F_{\bf{R},\lambda }(q)$ into smooth manifolds (see for instance
propositions 17 and 18 of the Chapter V in \cite{N}) and from the obvious
remark that when $\lambda $ belongs to the interior of a spectral band, then 
$F_{\bf{R},\lambda }(q)$ must separate $\bf{R}^n$. These two observations
imply existence of a smooth piece in $F_{\bf{R},\lambda }(q)$ of dimension
at least $(n-1)$.

Now we introduce our basic condition:

\begin{condition}
\label{condition}Assume that for any $\lambda $ that belongs to the interior
of a spectral band of the Schr\"{o}dinger operator 
\[
H_0=-\Delta +q(x)
\]
any irreducible component of the Fermi variety $F_{\,\lambda }(q)$
intersects the real space $\bf{R}^n$ by a subset of dimension $n-1$ (i.e.
by a subset that contains a piece of a smooth hypersurface).
\end{condition}

In fact, the Lemma \ref{big} says that for $\lambda $ in the interior of a
spectral band the Fermi variety $F_{\,\lambda }(q)$ does intersect the real
space $\bf{R}^n$ over a subset of dimension $n-1$. We, however, need more,
that every irreducible component of $F_{\,\lambda }(q)$ does the same. In
other words, there are no ``hidden'' components that do not show up on the
real subspace in any significant way. Thus, Lemma \ref{big} and
irreducibility of $F_{\,\lambda }(q)/2\pi \bf{Z}^n$ would imply validity of
the Condition \ref{condition}. In fact, we believe that (modulo the action
of the dual lattice) the Fermi surface is irreducible. The following
conjecture formulated in \cite{BKT} is probably correct:

\begin{conjecture}
\label{conj1}For any periodic potential (from an appropriate functional
class, for instance continuous, or locally square integrable) the surface $%
F_{\,\lambda }(q)/2\pi \bf{Z}^n$ is irreducible.
\end{conjecture}

It looks like this conjecture is very hard to prove (see related discussion
in \cite{BKT} and \cite{KT}). As the rest of the paper shows, proving it
would lead to a result on the absence of embedded eigenvalues. The following
weaker conjecture must be easier to prove:

\begin{conjecture}
\label{conj2}A generic periodic potential (from an appropriate functional
class) satisfies the Condition \ref{condition}.
\end{conjecture}

The word ``generic'' could mean ``from a residual set'', or something of
this sort.

One can easily prove using separation of variables and simple facts about
the Hill's equation irreducibility of $F_{\,\lambda }(q)/2\pi \bf{Z}^n$ for
separable periodic potentials $q(x)=\sum_iq_i(x_i)$ in any dimension \cite
{BKT}. The paper \cite{BKT}, however, also contains a stronger non-trivial
result:

\begin{lemma}
\label{irred} (\cite{BKT}) If $d=3$ and $q(x)=q_1(x_1)+q_2(x_2,x_3)$, then $%
F_{\,\lambda }(q)/2\pi \bf{Z}^n$ is irreducible.
\end{lemma}

The proof of this lemma relies on the deep study of the Bloch variety done
in \cite{KT}.

\section{The conditional result}

We are ready to prove our main conditional statement.

\begin{theorem}
\label{cond}If a real periodic potential $q(x)\in L^\infty (\bf{R}^n)$ ($%
n\leq $ $3$) satisfies the Condition \ref{condition} and an impurity
potential $v(x)$ is measurable and satisfies the estimate 
\begin{equation}
\left| \nu (x)\right| \leq Ce^{-\left| x\right| ^r},r>4/3
\label{estimate}
\end{equation}
almost everywhere in $\bf{R}^n$, then the spectrum of $H$ contains no
embedded eigenvalues. In other words, 
\[
\left\{ \lambda _j\right\} \cap \stackunder{i\geq 1}{\cup }%
(a_i,b_i)=\emptyset ,
\]
where $\left\{ \lambda _j\right\} $ is the impurity point spectrum of $H$,
and 
\[
\stackunder{i\geq 1}{\cup }[a_i,b_i]=\sigma (H_0)
\]
is the band structure of the essential spectrum of $H$.
\end{theorem}

\textbf{Proof}. Let us assume that there exists a $\lambda $ that belongs to
some $(a_i,b_i)$ and to the point spectrum of $H$ simultaneously. Then there
exists a non-zero function $u(x)\in L_2(\bf{R}^n)$ (an eigenfunction) such
that 
\[
-\Delta u+qu+vu=\lambda u, 
\]
or 
\[
(H_0-\lambda )u=-vu. 
\]
Let us denote the function in the right hand side by $\psi (x):$%
\[
\psi (x)=-v(x)u(x). 
\]
Consider the fundamental domain $\mathcal{K}$ of the group $\bf{Z}^n$ of
periods: 
\[
\mathcal{K}=\{(x_1,...,x_n)\in R^n|\;0\leq x_i\leq 1,\;i=1,...,n\}. 
\]
Then (\ref{estimate}) implies that the function $\psi (x)$ satisfies the
estimate 
\begin{equation}
\left| \left| \psi \right| \right| _{L^2(\mathcal{K}+l)}\leq Ce^{-\left|
l\right| ^r}  \label{decay}
\end{equation}
for all $l\in \bf{Z}^n$, where $r>4/3$. Consider the following Floquet
transform of functions defined on $\bf{R}^n$ (see section 2.2 in \cite{K}
about properties of this transform): 
\[
\mathcal{F}:\;f(x)\longrightarrow \widehat{f}(k,x)=\sum\limits_{l\in \bf{Z}%
^n}f(x-l)e^{-ik\cdot (x-l)}. 
\]
The result is a function, which is $\bf{Z}^n$-periodic with respect to $x$.
We can consider the resulting function as a function of $k$ with values in a
space of functions on the $n$-dimensional torus $\bf{T}^n=\bf{R}^n/\bf{Z}%
^n$. If we apply this transform to the function $\psi (x)$, then we get a
function of $k\in \bf{C}^n$ 
\begin{equation}
\phi (k)=\widehat{\psi }(k,x)  \label{funct}
\end{equation}
with values in the space $L^2(\bf{T}^n)$.

\begin{lemma}
\label{order}Function $\phi (k)$ is an entire function on $\bf{C}^n$ of the
order $s=r/(r-1)<4$.
\end{lemma}

\textbf{Proof of the lemma}. From (\ref{decay}) we get 
\[
\left| \left| \psi (x-l)e^{-ik\cdot (x-l)}\right| \right| _{L^2(\mathcal{K}%
)}\leq Ce^{C\left| k\right| +Im(k\cdot l)-\left| l\right| ^r}. 
\]
Then 
\[
\left| \left| \phi (k)\right| \right| _{L^2(\mathcal{K})}\leq Ce^{C\left|
k\right| }\stackunder{l\in \bf{Z}^n}{\sum }e^{-0.5\left| l\right| ^r}e^{%
Im(k\cdot l)-0.5\left| l\right| ^r}. 
\]
A simple Legendre transform type estimate (finding extremal values of the
exponent) shows that 
\[
e^{Im(k\cdot l)-0.5\left| l\right| ^r}\leq Ce^{C\left| k\right|
^{r/(r-1)},} 
\]
which proves that the function $\phi (k)$ is an entire function of the order 
$s=r/(r-1)$ in $\bf{C}^n$. The lemma is proven.

As it is well known (see e.g. Chapter 2 of \cite{K} or XIII in \cite{RS}),
the transform $\mathcal{F}$ leads to the following operator equation with a
parameter $k$: 
\begin{equation}
\left( H_0(k)-\lambda \right) \widehat{u}(k)=\phi (k),  \label{operator}
\end{equation}
where 
\[
\widehat{u}(k)=\left( \mathcal{F}u\right) (k,x) 
\]
is defined for $k\in \bf{R}^n$ and 
\[
H_0(k)=(i\nabla -k)^2+q(x) 
\]
is considered as an operator on the torus $\bf{T}^n$. According to Theorem
2.2.5 in \cite{K}, 
\[
\widehat{u}(k)\in L_{loc}^2(\bf{R}^n,L_2(\bf{T}^n)). 
\]
In fact, standard interior elliptic estimates show that 
\[
\widehat{u}(k)\in L_{loc}^2(\bf{R}^n,H^2(\bf{T}^n)), 
\]
where $H^2(\bf{T}^n)$ is the Sobolev space of order two on $\bf{T}^n$.
Theorem 3.1.5 of \cite{K} implies that the operator 
\[
\left( H_0(k)-\lambda \right) :H^2(\bf{T}^n)\rightarrow L_2(\bf{T}^n) 
\]
is invertible if and only if $k\notin F_{\bf{R},\lambda }(q)$. As it is
shown in the proof of Theorem 3.3.1 and in Lemma 1.2.21 of \cite{K} (see
also comments in Section 3.4.D on reducing requirements on the potential, in
particular Theorem 3.4.2), the inverse operator can be represented as
follows: 
\begin{equation}
\left( H_0(k)-\lambda \right) ^{-1}=B(k)/\zeta (k),  \label{represent}
\end{equation}
where $B(k)$ is a bounded operator from $L_2(\bf{T}^n)$ into $H^2(\bf{T}%
^n) $ and $B(k)$ and $\zeta (k)$ are correspondingly an operator and a
scalar entire functions of order $n$ in $\bf{C}^n$. Besides, the zeros of $%
\zeta (k)$ constitute exactly the Fermi variety $F_{\,\lambda }(q)$. We
conclude now that the following representation holds on the set $\bf{R}%
^n\backslash F_{\bf{R},\lambda }(q)$: 
\[
\widehat{u}(k)=\frac{B(k)\phi (k)}{\zeta (k)}=\frac{g(k)}{\zeta (k)}, 
\]
where $g(k)$ is a $H^2(\bf{T}^n)$-valued entire function of order $n$.

Now we need the following auxiliary result:

\begin{lemma}
\label{divis}Let $Z$ be the set of all zeros of an entire function $\zeta (k)
$ in $\bf{C}^n$ and $Z_j$ be its irreducible components. Assume that the
real part $Z_{j,\bf{R}}=Z_j\cap \bf{R}^n$ of each $Z_j$ contains a
submanifold of real dimension $n-1$. Let also $g(k)$ be an entire function
in $\bf{C}^n$ with values in a Hilbert space $H$ such that on the real
subspace $\bf{R}^n$ the ratio 
\[
\widehat{u}(k)=\frac{g(k)}{\zeta (k)}
\]
belongs to $L_{2,loc}(\bf{R}^n,H)$. Then $\widehat{u}(k)$ extends to an
entire function with values in $H$.
\end{lemma}

\textbf{Proof of the lemma}. Applying linear functionals, one can reduce the
problem to the case of scalar functions $g$, so we will assume that $g(k)\in 
\bf{C}$.

According to Lemma \ref{property}, the sets of regular points of components $%
Z_j$ are disjoint. Hence, the traces of these sets on $\bf{R}^n$ are also
disjoint. Consider one component $Z_j$. The intersection of $regZ_j$ with
the real space $\bf{R}^n$ contains a smooth manifold of dimension $(n-1)$.
Namely, we know that $Z_{j,\bf{R}}$ contains such a manifold, which we will
denote $M_j$. The only alternative to our conclusion would be that the whole 
$M_j$ sits inside the singular set of $Z_j$. The proof of Corollary \ref{cor}
shows that this is impossible, since lower dimensional strata cannot contain
any open pieces of $M_j$.

Let us denote by $m_j$ the minimal order of zero of function $\zeta (k)$ on $%
Z_j$. (We remind the reader that the order of zero of an analytic function
at a point is determined by the order of the first non-zero term of the
function's expansion at this point into homogeneous polynomials.) Since the
condition that an analytic function has a zero of order higher than a given
number can be written down as a finite number of analytic equations, one can
conclude that the order of zero of $\zeta (k)$ equals $m_j$ on a dense open
subset of $Z_j$, whose complement is an analytic subset of lower dimension.
As it was explained in the proof of Corollary \ref{cor}, lower dimensional
strata cannot contain $(n-1)$-dimensional submanifolds of $\bf{R}^n$. This
means that one can find a point $k^j\in Z_{j,\bf{R}}$ such that $Z_{j,\bf{R%
}}$ is a smooth hypersurface in a neighborhood $U$ of $k^j$ in $\bf{R}^n$
and such that the order of zero of $\zeta (k)$ on $Z_{j,\bf{R}}$ equals $%
m_j $ in this neighborhood. Let us now prove that $g(k)$ has zeros on $Z_{j,%
\bf{R}}\cap U$ of at least the same order as $\zeta (k)$. If this were not
so, then in appropriate local coordinates $k=(k_1,...,k_n)$ the ratio $%
g/\zeta $ would have a singularity of at least the order $k_1^{-1}$, which
implies local square non-integrability of the function $\widehat{u}(k)$.
This contradicts our assumption, so we conclude that $g(k)$ has zeros on $%
Z_{j,\bf{R}}\cap U$ of at least the same order as $\zeta (k)$. Consider the
(analytic) set $A$ of all points in $\bf{C}^n$ where $g(k)$ has zeros of
order at least $m_j$. We have just proven that the intersection of $A$ with $%
Z_j$ contains an $(n-1)$-dimensional smooth submanifold of $R^n$. Then
Corollary \ref{cor} implies that $Z_j\subset A$, or $g(k)$ has zeros on $Z_j$
of at least the order $m_j$. Hence, according to Proposition 3 in section
1.5 of Chapter 1 in \cite{Ch} the ratio $g/\zeta $ is analytic everywhere in 
$\bf{C}^n$ except maybe at the union of subsets of $Z_j$, where the
function $\zeta $ has zeros of order higher than $m_j$. This subset,
however, is of dimension not higher than $n-2$. Now a standard analytic
continuation theorem (see for instance Proposition 3 in Section 1.3 of the
Appendix in \cite{Ch}) guarantees that $g/\zeta $ is an entire function.
This concludes the proof of the lemma.

Returning to the proof of the theorem and using the result of the Lemma \ref
{big} and the assumption that the Condition \ref{condition} is satisfied for
the potential $q(x)$, we conclude that $\widehat{u}(k)$ is an entire
function and that it is the ratio of two entire functions of order at most $%
w=\max (n,s)<4$, where $s$ is defined in Lemma \ref{order}. Using (in the
radial directions) the estimate of entire functions from below contained in
section 8 of Chapter 1 in \cite{Le} (see also Theorem 1.5.6 and Corollary
1.5.7 in \cite{K} or similar results in \cite{Bo}), we conclude that $%
\widehat{u}(k)$ is itself an entire function of order $w$ with values in $%
H^2(\bf{T}^n)$. Now Theorem 2.2.2 of \cite{K} claims that the solution $u(x)
$ satisfies the decay estimate: 
\[
\left| \left| u\right| \right| _{H^2(K+a)}\leq C_p\exp (-c_p\left| a\right|
^p)
\]
for any $p<w/(w-1)$, where $K$ is an arbitrary compact in $\bf{R}^n$, and $%
a\in \bf{R}^n$. Using standard embedding theorems, we conclude that 
\[
\left| u(x)\right| \leq C_p\exp (-c_p\left| x\right| ^p),\;\forall
\;p<w/(w-1).
\]
Since $w<4$, one can choose a value $p$ such that 
\[
4/3<p<w/(w-1).
\]
However, Remark 2.6 in \cite{FH} and Theorem 1 in \cite{M} state that there
is no non-trivial solution of the equation 
\[
-\Delta u+qu+vu-\lambda u=0
\]
with the rate of decay 
\[
\left| u(x)\right| \leq C\exp (-c\left| x\right| ^p),\;p>4/3.
\]
This contradiction concludes the proof of the theorem.%

\begin{remark}
The condition $p>4/3$ (and hence the dimension restriction $n<4$) is
essential for the validity of the result of \cite{FH} and \cite{M}, so this
is the place where our argument breaks down for dimensions four and higher
even if we require compactness of support of the perturbation potential $v(x)
$. The rest of the arguments stay intact (the embedding theorem argument,
which also depends on dimension, is not really necessary).
\end{remark}

\section{Separable potentials}

The result of the previous section leads to the problem of finding classes
of periodic potentials that satisfy the Condition \ref{condition}. As we
stated in conjectures \ref{conj1} and \ref{conj2}, we believe that all (or
almost all) of periodic potentials satisfy this condition. Although we were
not able to prove these conjectures, as an immediate corollary of the Lemma 
\ref{irred} and of the Theorems \ref{cond} and \ref{sep} we get the
following result:

\begin{theorem}
\label{combin}If for $n<4$ the background periodic potential $q(x)\in
L^\infty (\bf{R}^n)$ is separable for $n=2$ or $q(x)=q_1(x_1)+q_2(x_2,x_3)$
for $n=3$, and the perturbation potential $v(x)$ satisfies the estimate \ref
{estimate}, then there are no eigenvalues of the operator $H$ in the
interior of the bands of the continuous spectrum.
\end{theorem}

\section{Comments}

1. We have only proven the absence of eigenvalues embedded into the interior
of a spectral band. It is likely that eigenvalues cannot occur at the ends
of the bands either (maybe except the bottom of the spectrum), if the
perturbation potential decays fast enough. This was shown in the
one-dimensional case in \cite{RB} under the condition 
\[
\int (1+\left| x\right| )\left| v(x)\right| dx<\infty 
\]
on the perturbation potential. On the other hand, if $v(x)$ only belongs to $%
L^1$, then the eigenvalues at the endpoints of spectral bands can occur \cite
{RB2}.

2. Most of the proof of the conditional Theorem \ref{cond} does not require
the unperturbed operator to be a Schr\"{o}dinger operator. One can treat
general selfadjoint periodic elliptic operators as well. The only obstacle
occurs at the last step, when one needs to conclude the absence of fast
decaying solutions to the equation. Here we applied the results of \cite{FH}
and \cite{M}, which are applicable only to the operators of the
Schr\"{o}dinger type. Carrying over these results to more general operators
would automatically generalize Theorem \ref{cond}. The restriction that the
dimension $n$ is less than four also comes from the allowed rate of decay
stated in the result of \cite{FH} and \cite{M}.

3. It might seem that the irreducibility condition \ref{condition} arises
only due to the way the proof is done. We believe that this is not true, and
that the validity of the condition is essentially equivalent to the absence
of embedded eigenvalues. To be more precise, we conjecture that existence
for some $\lambda $ in the interior of a spectral band of an irreducible
component $A$ of the Fermi surface such that $A\cap \bf{R}^n=\emptyset $
implies existence of a localized perturbation of the operator that creates
an eigenvalue at $\lambda $. As a supporting evidence of this one can
consider fourth order periodic differential operators, where the Fermi
surface contains four points. In this case one can have $\lambda $ in the
continuous spectrum, while some points of the Fermi surface being complex.
Then one can use these components of the Fermi surface ``hidden'' in the
complex domain to cook up a localized perturbation that does create an
eigenvalue at $\lambda $ \cite{P}.

\begin{center}
ACKNOWLEDGMENTS
\end{center}

The authors express their gratitude to Professors A. Figotin, T.
Hoffmann-Ostenhof, H. Kn\"{o}rrer, S. Molchanov, and V. Papanicolaou for
helpful discussions and information. The work of P. Kuchment was partly
supported by the NSF Grant DMS 9610444 and by a DEPSCoR Grant. P. Kuchment
expresses his gratitude to NSF, ARO, and to the State of Kansas for this
support. The work of B. Vainberg was partly supported by the NSF Grant
DMS-9623727. B. Vainberg expresses his gratitude to NSF for this support.
The content of this paper does not necessarily reflect the position or the
policy of the federal government, and no official endorsement should be
inferred.


\begin{thebibliography}{99}
\bibitem{AADH}  S. Alama, M. Avellaneda, P. A. Deift, and R. Hempel, On the
existence of eigenvalues of a divergence form operator $A+\lambda B$ in a
gap of $\sigma (A)$, Asymptotic Anal. 8(1994), no.4, 311-314.

\bibitem{ADH}  S. Alama, P. A. Deift, and R. Hempel, Eigenvalue branches of
the Schr\"{o}dinger operator $H-\lambda W$ in a gap of $\sigma (H)$, Commun.
Math. Phys. 121(1989), 291-321.

\bibitem{BKT}  D. B\"{a}ttig, H. Kn\"{o}rrer, and E. Trubowitz, A
directional compactification of the complex Fermi surface, Compositio Math.
79(1991), no.2, 205-229.

\bibitem{B}  M. S. Birman, The discrete spectrum of the periodic
Schr\"{o}dinger operator perturbed by a decreasing potential, Algebra i
Analiz 8(1996), no.1, 3-20.

\bibitem{B2}  M. S. Birman, On the spectrum of singular boundary value
problems, Mat. Sbornik 55(1961), no.2, 125-173.

\bibitem{B3}  M. S. Birman, On the discrete spectrum in the gaps of a
perturbed periodic second order operator, Funct. Anal. Appl. 25(1991),
158-161.

\bibitem{Bo}  R. Boas, Jr., Entire Functions, Acad. Press, NY 1954.

\bibitem{Ch}  E. M. Chirka, Complex analytic sets, Mathematics and its
Applications (Soviet Series), 46. Kluwer Academic Publishers Group,
Dordrecht, 1989.

\bibitem{D}  P. A. Deift and R. Hempel, On the existence of eigenvalues of
the Schr\"{o}dinger operator $H-\lambda W$ in a gap of $\sigma (H)$, Commun.
Math. Phys. 103(1986), 461-490.

\bibitem{E}  M. S. P. Eastham, The Spectral Theory of Periodic Differential
Equations, Scottish Acad. Press, Edinburgh - London 1973.

\bibitem{EK}  M. S. P. Eastham and H. Kalf, Schr\"{o}dinger-type Operators
with Continuous Spectra, Pitman. Boston 1982.

\bibitem{FH}  R. Froese, I. Herbst, M. Hoffmann-Ostenhof, and T.
Hoffmann-Ostenhof, $L^2$-lower bounds to solutions of one-body
Schr\"{o}dinger equations, Proc. Royal Soc. Edinburgh, 95A(1983), 25-38.

\bibitem{Gelf}  I.M. Gelfand, Expansion in eigenfunctions of an equation
with periodic coefficients, Dokl. Akad. Nauk SSSR 73(1950), 1117-1120.

\bibitem{GS}  F. Gesztesy and B. Simon, On a theorem of Deift and Hempel,
Commun. Math. Phys. 116(1988), 503-505.

\bibitem{G}  I. M. Glazman, Direct Methods of Qualitative Spectral Analysis
of Singular Differential Operators, I.P.S.T., Jerusalem 1965.

\bibitem{GR}  R. C. Gunning and H. Rossi, Analytic Functions of Several
Complex Variables, Prentice-Hall, Englewood Cliffs, NJ 1965.

\bibitem{He}  R. Hempel, Eigenvalue branches of the Schr\"{o}dinger operator 
$H\pm \lambda W$ in a spectral gap of $H$, J. Reine Angew. Math. 399(1989),
38-59.

\bibitem{H}  L. H\"{o}rmander, The Analysis of Linear Partial Differential
Operators, v. 2, Springer Verlag, Berlin 1983.

\bibitem{KT}  H. Kn\"{o}rrer and J. Trubowitz, A directional
compactification of the complex Bloch variety, Comment. Math. Helv.
65(1990), 114-149.

\bibitem{K}  P. Kuchment, Floquet Theory for Partial Differential Equations,
Birkh\"{a}user, Basel 1993.

\bibitem{KV}  P. Kuchment and B. Vainberg, On embedded eigenvalues of
perturbed periodic Schr\"{o}dinger operators, in Spectral and scattering
theory (Newark, DE, 1997), 67--75, Plenum, New York,1998.

\bibitem{Lel}  P. Lelong and L. Gruman, Entire Functions of Several Complex
Variables, Springer Verlag, Berlin 1986.

\bibitem{L}  S. Z. Levendorskii, Asymptotic formulas with remainder
estimates for eigenvalue branches of the Schr\"{o}dinger operator $H-\lambda
W$ in a gap of $H$. To appear in Transactions of American Mathematical
Society.

\bibitem{L2}  S. Z. Levendorskii and S. I. Boyarchenko, An asymptotic
formula for the number of eigenvalue branches of a divergence form operator $%
A+\lambda B$ in a spectral gap of $A$. To appear in Communications in Part.
Differ. Equat.

\bibitem{L3}  S. Z. Levendorskii, Lower bounds for the number of eigenvalue
branches for the Schr\"{o}dinger operator $H-\lambda W$ in a gap of $H$: the
case of indefinite $W$, Comm. partial Diff. Equat. 20(1995), no.5-6, 827-854.

\bibitem{Le}  B. Levin, Distribution of Zeros of Entire Functions, Transl.
Math. Monogr. v.5, Amer. Math. Soc., Providence, RI 1964.

\bibitem{M}  V. Meshkov, On the possible rate of decay at infinity of
solutions of second order partial differential equations, Mat. Sbornik,
182(1991), no.3, 364-383. English translation in Math. USSR Sbornik
72(1992), no.2, 343-351.

\bibitem{N}  R. Narasimhan, Intorduction to the Theory of Analytic Spaces,
Lect. Notes in Math., v. 25, Springer Verlag, Berlin 1966.

\bibitem{OK}  F. Odeh, J.B. Keller, Partial differential equations with
periodic coefficients and Bloch waves in crystals, J. Math. Phys. 5(1964),
1499-1504.

\bibitem{P}  V. Papanicolaou, private communication, 1998.

\bibitem{Ra}  G. D. Raikov, Eigenvalue asymptotics for the Schr\"{o}dinger
operator with perturbed periodic potential, Invent. Math. 110(1992), 75-93.

\bibitem{RS}  M. Reed and B. Simon, Methods of Modern Mathematical Physics
v. 4, Acad. Press, NY 1978.

\bibitem{RB}  F. S. Rofe-Beketov, A test for the finiteness of the number of
discrete levels introduced into the gaps of a continuous spectrum by
perturbations of a periodic potential, Soviet Math. Dokl. 5(1964), 689-692.

\bibitem{RB2}  F. S. Rofe-Beketov, Spectrum perturbations, the Knezer-type
constants and the effective mass of zones-type potentials, in ``Constructive
Theory of Functions'84'', Sofia 1984, p.757-766.

\bibitem{S}  A. V. Sobolev, Weyl asymptotics for the discrete spectrum of
the perturbed Hill operator, Adv. Sov. Math. 7(1991), 159-178.

\bibitem{T}  E.C. Titchmarsh, Eigenfunction Expansions Associated with
Second-Order Differential Equations, Part II, Claredon Press, Oxford 1958.

\bibitem{V}  B. Vainberg, Principles of radiation, limiting absorption and
limiting amplitude in the general theory of partial differential equations,
Russian Math. Surveys 21(1966), no 3, 115-193.

\bibitem{W}  C. Wilcox, Theory of Bloch waves, J. Analyse Math. 33(1978),
146-167.
\end{thebibliography}
\end{document}